\documentclass[aps,prd,twocolumn,10pt]{revtex4-2}

\usepackage{amsmath,amssymb}
\usepackage{booktabs}
\usepackage{graphicx}
\usepackage[colorlinks=true,allcolors=blue]{hyperref}
\usepackage[normalem]{ulem}
\usepackage{xcolor}

\usepackage{xcolor}
\usepackage{soul}

\setstcolor{red}  
\begin{document}

\title{Axion-Like Particle Dark Matter Intensity Mapping: A New Probe via Cross-Correlation with Galaxy Surveys}

\author{Wen-Qing Guo}
\email{wetingguo@gmail.com}
\affiliation{Department of Physics, Stellenbosch University, Matieland 7602, South Africa}

\begin{abstract}
The particle nature of dark matter (DM) remains one of the most significant enigmas in modern cosmology. Axion-like particles (ALPs), as well-motivated candidates for cold dark matter, can undergo radiative decay into photon pairs, a process that is significantly enhanced in the presence of ambient radiation fields. In this work, we propose a novel probe of $\mu{\rm eV}$-scale ALP DM by cross-correlating radio intensity mapping (IM) with the large-scale galaxy distribution from the 2MASS Redshift Survey (2MRS) in the local universe ($z\leq 0.1$). We develop a comprehensive theoretical framework that incorporates stimulated decay effects driven by both the Cosmic Microwave Background (CMB) and a bottom-up modeled extragalactic radio background (ERB). By forecasting the sensitivity of the Square Kilometre Array (SKA) Phase 2, we demonstrate that this cross-correlation technique provides a promising and complementary approach to searching for ALP DM signals. This study establishes a new proof-of-concept for utilizing next-generation radio telescopes to probe ALP dark matter on cosmic scales.
\end{abstract}

\maketitle
\section{Introduction}

The existence of dark matter (DM) is robustly supported by a wealth of astrophysical and cosmological observations. Within the concordance $\Lambda$CDM framework, cosmic microwave background (CMB) measurements indicate that DM accounts for approximately 25\% of the universe's total energy density \citep{Planck:2018vyg}. Given its fundamental role in the evolution of cosmic structures, unraveling the particle nature of DM remains a pivotal pursuit in modern physics.

To facilitate efficient structure formation, DM must be sufficiently cold and stable over cosmological timescales, possessing only extremely weak interactions with Standard Model (SM) particles \citep{Blumenthal:1984bp}. Weakly Interacting Slim Particles (WISPs), characterized by their sub-eV masses, emerge as compelling candidates that satisfy these criteria \citep{Ringwald:2012hr, Arias:2012az}. Among these, axion-like particles (ALPs) are particularly well motivated~\citep{Arias:2012az, Marsh:2015xka, Irastorza:2018dyq, Co:2020xlh}, often appearing as natural extensions of the SM in string-inspired theories \citep{Irastorza:2018dyq}. In addition to being thermally produced as light DM, ALPs can also be generated non-thermally via the misalignment mechanism in the early universe, spanning a broad mass range of $10^{-20}\,\rm eV \leq m_{a} \leq {\rm eV}$~\citep{Preskill:1982cy, Abbott:1982af, Dine:1982ah, Adams:2022pbo}.

Crucially, ALPs are generally decoupled from the Peccei-Quinn (PQ) mechanism \citep{Irastorza:2021tdu}, 
thereby rendering the coupling constant independent of the particle mass. The coupling of ALPs to photons facilitates their conversion into photons via the Primakoff effect in external magnetic fields, as well as their radiative decay into photon pairs. These processes underpin the search for ALPs in laboratory experiments \citep{Bahre:2013ywa, ADMX:2019uok} and astrophysical observations of DM-rich systems \citep{Fermi-LAT:2016nkz, CAST:2017uph, Xia:2019yud, Li:2020pcn, Caputo:2018vmy}, most notably around neutron stars \citep{Hook:2018iia, Foster:2022fxn, Li:2025zgp} and within dwarf spheroidal galaxies (dSphs) \citep{Caputo:2018ljp, Guo:2024oqo}. However, these studies predominantly probe relatively small angular scales, constrained by the local environments of specific DM halos.

To overcome these limitations, the search for ALP DM can be extended to cosmological scales by statistically correlating the spatial distribution of DM tracers with the expected ALP-induced emission. A promising technique for such large-scale searches is intensity mapping (IM). Originally developed for neutral hydrogen (HI), CO, and other molecular lines, IM captures the aggregate emission from unresolved sources over a wide range of redshifts, effectively mapping the three-dimensional large-scale structure (LSS) of the Universe. {The application of IM to ALP dark matter has been explored in Refs.~\cite{Bauer:2020zsj, Shirasaki:2021yrp, Wu:2025uvi}}. Since the radiative decay of ALP DM into two photons produces a narrow spectral line, the resulting signal manifests as a cosmological line emission, analogous to the HI 21 cm signal. Consequently, the IM framework, particularly when cross-correlated with galaxy or galaxy cluster surveys, provides a powerful method to extract faint ALP-induced signals from complex astrophysical foregrounds.

In this work, we assess the feasibility of employing ALP IM as a novel probe of ALP DM decay. Specifically, we forecast the sensitivity of future radio observations, exemplified by the Square Kilometre Array (SKA), in detecting ALP DM signals through the cross-correlation between ALP-induced line emissions and large-scale galaxy distributions. By modeling the noise properties of the SKA Phase 2 (SKA2), we provide projections for the ALP-photon coupling parameter $g_{a\gamma\gamma}$. Our work serves as a proof-of-concept study, aiming to inspire innovative methodologies in the search for ALP DM.

The paper is organized as follows. In Section \ref{Sec:Theore}, we delineate the theoretical framework for the angular power spectrum. In Section \ref{Sec:ResultDiscussion}, we present and evaluate our forecast results. Finally, we conclude in Section \ref{Sec:Conclusion}. Throughout this paper, we adopt the spatially flat $\Lambda$CDM cosmological model with {\it Planck}-2018 cosmological parameters: $H_{0} = 67.32\ \mathrm{km\,s^{-1}\,Mpc^{-1}}$, $\Omega_{\rm c}h^{2} = 0.1201$, and $\Omega_{\rm b}h^{2} = 0.02238$ \citep{Planck:2018vyg}. Furthermore, we utilize natural units, namely $k_{B}=c=\hbar=1$, in specific contexts, with explicit notification provided upon their application.

\section{The cross-correlation signal}\label{Sec:Theore}
\subsection{Angular power spectrum}
The angular power spectrum (APS) of the cross-correlation between two observables $i$ and $j$ is given by \citep{Fornengo:2013rga}
\begin{equation}\label{APS}
    C_{\ell}^{ij}=\int \frac{{\rm d}\chi}{\chi^2}W_{i}(\chi)W_{j}(\chi)P_{ij}\left(k=\frac{\ell+1/2}{\chi},z \right),
\end{equation}
where $\ell$ is the multipole and $\chi(z)$ is the comoving distance, with ${\rm d}\chi = c\,{\rm d}z/H(z)$. The Hubble parameter for a flat $\Lambda$CDM cosmology is $H(z) = H_{0}\sqrt{\Omega_{\rm m}(1+z)^{3} + \Omega_{\Lambda}}$, where $\Omega_{\Lambda}=1-\Omega_{\rm m}$ and $\Omega_{\rm m}=0.3158$. 
Throughout this work, $\chi$ and redshift $z$ are used interchangeably. 
The window function $W(\chi)$ describes the redshift-dependent contribution of the observable to the cross-correlation signal. 

The quantity $P_{ij}$ represents the three-dimensional cross-power spectrum between the fluctuations of two fields $g_{i}$ and $g_{j}$, where the fluctuation field is defined as $f(\chi, \vec{\theta}) = g(\chi, \vec{\theta}) - \langle g(\chi, \vec{\theta})\rangle$, and $\langle\cdots \rangle$ denotes the ensemble average~\citep{Fornengo:2013rga}. 
Within the halo model framework~\citep{Cooray:2002dia}, $P_{ij}$ is decomposed into one-halo and two-halo terms, namely $P_{ij}=P_{ij}^{1\rm h} + P^{2\rm h}_{ij}$, which are expressed as
\begin{equation}
\begin{aligned}
P^{1\rm h}_{ij}&(k,z)=\int {\rm d}M\frac{{\rm d}n}{{\rm d}M}(M,z)\frac{\tilde{g}^*_i(k|M,z)}{\langle{g}_i\rangle(z)}\frac{\tilde{g}_j(k|M,z)}{\langle{g}_j\rangle(z)},\\
P^{2\rm h}_{ij}&(k,z)=\left[\int {\rm d}M_1\frac{{\rm d}n}{{\rm d}M_1}(M_1,z)b_{i}(M_1,z)\frac{\tilde{g}^*_{i}(k|M_1,z)}{\langle{g}_i\rangle(z)}\right]\times\\
&\left[\int {\rm d}M_2\frac{{\rm d}n}{{\rm d}M_2}(M_2,z)b_{j}(M_2,z)\frac{\tilde{g}_{j}(k|M_2,z)}{\langle{g}_j\rangle(z)}\right]P_{\rm lin}(k,z).
\end{aligned}
\end{equation}
Here $M$ denotes the halo mass, and the integral is performed over the range 
$M_{\rm min}=10^{-6}\,{\rm M_\odot}$ to $M_{\rm max}=10^{18}\,{\rm M_\odot}$. 
The halo mass function ${\rm d}n/{\rm d}M$ follows the ``Sheth-Tormen'' form~\citep{Sheth:1999su}. 
The terms $b_i(M,z)$ and $b_j(M,z)$ are the linear bias factors relating observables $i$ and $j$ to the underlying matter density field. 
$\tilde{g}(k|M,z)$ is the Fourier transform of the field profile ${g}(r|M,z)$ for a halo of mass $M$ at redshift $z$, and the asterisk denotes complex conjugation. 
The linear matter power spectrum $P_{\rm lin}(k,z)$ is computed using the public code $\tt CAMB$\footnote{\url{https://camb.readthedocs.io/en/latest/}}. 

In this work, we specifically focus on the cross-correlation between radio emission from ALP dark matter decay and a galaxy catalog. 
\subsection{ALP DM decay}

In vacuum, an ALP of mass $m_{a}$ can undergo spontaneous decay into two photons. Energy conservation implies that each photon carries an energy $E_{\gamma}=m_{a}/2$, corresponding to a photon frequency $\nu = m_{a}/4\pi$ in natural units. The spontaneous decay lifetime $\tau_{a}$ is determined by the ALP mass and the ALP–photon coupling parameter $g_{a\gamma\gamma}$ 
\begin{equation}
    \tau_{a} = \frac{64\pi}{m_{a}^{3}g_{a\gamma\gamma}^{2}}.
\end{equation}

\begin{table}[t]
    \centering
    \caption{Best-fit parameters for the stellar–halo mass relation (SHMR) adopted in this work, as reported in Ref.~\cite{Girelli:2020goz}.}
    \label{SHMR_table}
    \addtolength{\tabcolsep}{1pt} 
    \begin{tabular}{l ccc cccc c}
        \toprule
        \toprule 
         & $C$ & $\phi$ & $D$ & $\mu$ & $E$ & $F$ & $G$ & $\eta$ \\
        \midrule
        Best fit & 0.046 & $-0.38$ & 11.79 & 0.20 & 0.043 & 0.96 & 0.709 & $-0.18$ \\
        \bottomrule
        \bottomrule
    \end{tabular}
\end{table}

For the parameter space relevant to ALP DM, these particles possess extremely long lifetimes, typically far exceeding the age of the Universe. Consequently, detecting radio signals originating from the spontaneous decay of $\mu{\rm eV}$-scale ALPs is highly challenging for current radio telescopes. However, the decay rate can be significantly enhanced through stimulated emission when the decay occurs in the presence of an ambient radiation bath \cite{Caputo:2018vmy}. Quantitatively, the effective decay rate is multiplied by a factor $(1+2f_{\gamma})$, where $f_{\gamma}$ denotes the photon occupation number of the ambient radiation field at the relevant frequency. For $\mu{\rm eV}$-scale ALPs, the primary contributors at radio frequencies are the CMB and the Extragalactic Radio Background (ERB). The CMB follows a blackbody spectrum, yielding a photon occupation number
\begin{equation}
    f_{\gamma,{\rm CMB}}(z)=\frac{1}{e^{h\nu/k_{\rm B}T_{\rm CMB}(z)} - 1},
\end{equation}
where $T_{\rm CMB}(z) = 2.73(1+z)\,{\rm K}$.

To evaluate the ERB contribution, we adopt a bottom-up approach by integrating the radio emission from individual galaxies over cosmological volumes, {thereby modeling the isotropic ERB component and neglecting its spatial fluctuations}. The radio luminosity of a galaxy at $150\,{\rm MHz}$ is empirically scaled with its Star Formation Rate (SFR) {and its stellar mass $M_{*}$} \cite{2018MNRAS.475.3010G}
{
\begin{equation}\label{Re2:eqSFR1}
    \left(\frac{L_{\rm 150\,MHz}}{\rm W\,Hz^{-1}}\right) = 10^{22.13}\left(\frac{\rm SFR}{\rm M_{\odot}yr^{-1}}\right)^{0.77}\left(\frac{M_{*}}{10^{10}\,{\rm M_{\odot}}}\right)^{0.43}.
\end{equation}
}

The SFR itself is also connected to the stellar mass through the star-forming main sequence, which evolves with cosmic time \cite{2014ApJS..214...15S}
\begin{equation}
\begin{aligned}
    \log_{10} {\rm SFR}(\rm M_{\odot}\,yr^{-1}) & = (0.84 - 0.026t)\log_{10}\left(M_{*} / {\rm M_{\odot}}\right) \\
    &\quad- (6.51 - 0.11t),
\end{aligned}
\end{equation}
where $t$ denotes the cosmic age in units of $\rm Gyr$.

To relate galaxy properties to the underlying dark matter halos, we adopt a Stellar–Halo Mass Relation (SHMR) that relates the stellar mass to the host halo mass $M$ \cite{Girelli:2020goz}
\begin{equation}
    \frac{M_{*}}{M}(z) = 2A(z)\left[\left(\frac{M}{M_{A}(z)}\right)^{-\beta(z)} + \left(\frac{M}{M_{A}(z)}\right)^{\gamma(z)}\right]^{-1},
\end{equation}
where the redshift-dependent quantities are defined as
\begin{equation}
\begin{aligned}
    & A(z) = C(1+z)^{\phi} ,\\
    & \log_{10}M_{A}(z) = D +z\cdot\mu , \\
    & \beta(z) = E\cdot z + F, \\
    &\gamma(z) = G(1+z)^{\eta}.
\end{aligned}
\end{equation}
The parameter values adopted in this work are listed in Table~\ref{SHMR_table}.

The specific radio luminosity $L_{\nu}$ at a frequency $\nu$ is then extrapolated using a typical power-law spectral index \cite{2018MNRAS.475.3010G}
\begin{equation}
    L_{\nu} = L_{\rm 150\,MHz}\left(\frac{\nu}{150\,{\rm MHz}}\right)^{-0.7}.
\end{equation}

To determine the total radiation field, we estimate the volume emissivity $j_{\nu}(z)$ by convolving the halo mass function with the halo-dependent luminosity
\begin{equation}
    j_{\nu}(z) = \int{\rm d}M \frac{{\rm d}n}{{\rm d}M} L_{\nu}(M,z).
\end{equation}

\begin{figure}
    \centering
    \includegraphics[width=1.\linewidth]{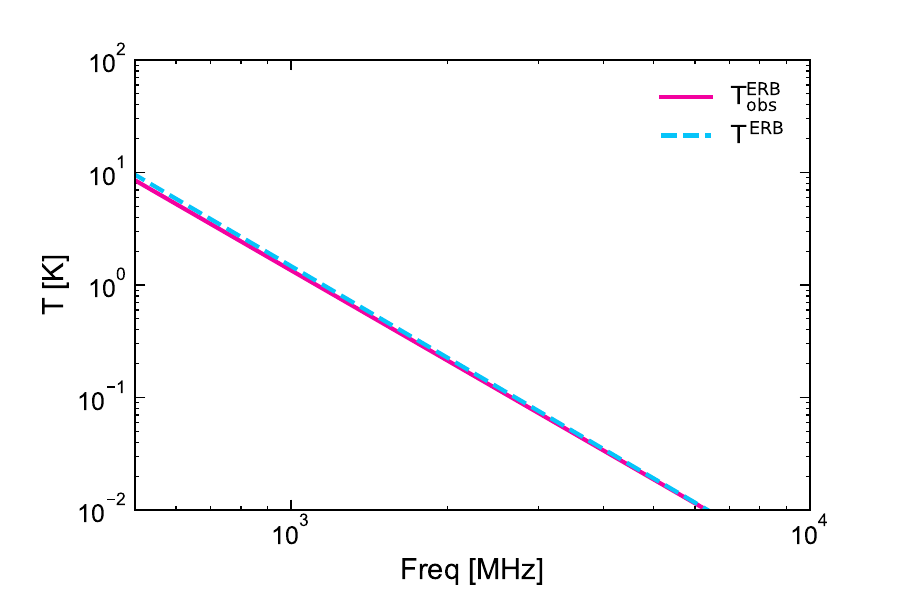}
    \caption{Comparison between the predicted ERB brightness temperature at $z=0$ (dashed blue) and the observed extragalactic radio background temperature from ARCADE2 (solid pink)~\cite{Fixsen2011, Singal:2022jaf}.}
    \label{fig:ERB_Tb}
\end{figure}

Accounting for cosmological propagation and redshifting of photon energies, the specific intensity of the ERB $I^{\rm ERB}_{\nu}(z)$ of  at redshift $z$ is obtained by integrating contributions over the past light cone $z'>z$
\begin{equation}
    I^{\rm ERB}_{\nu}(z) = \frac{c}{4\pi}\int_{z}^{\infty} \frac{{\rm d} z'}{(1+z')H(z')}j_{\nu'}(z'),
\end{equation}
where $\nu' = \nu (1+z')/(1+z)$. { The corresponding brightness temperature can be calculated in the Rayleigh-Jeans approximation as
\begin{equation}
    T^{\rm ERB}_{\nu}(z) = \frac{I^{\rm ERB}_{\nu}(z)c^{2}}{2k_{\rm B}\nu^{2}}.
\end{equation}

The ERB has been well measured to exhibit a frequency-dependent temperature~\cite{Fixsen2011, Fornengo:2014mna, Singal:2022jaf}. A comparison between our model at $z=0$ (dashed blue) and the observed ERB temperature $T^{\rm ERB}_{\rm obs}$ (solid pink), corresponding to the present-day ($z=0$) background, is shown in Fig.~\ref{fig:ERB_Tb}. We find that our model
{has successfully reproduced both the amplitude and the spectral
features of the observed ERB.}

{ Finally, the photon occupation number associated with the ERB at redshift $z$ is
then
\begin{equation}
    f_{\gamma,{\rm ERB}}(z) = \frac{c^{2}}{2h\nu^{3}}I^{\rm ERB}_{\nu}(z).
\end{equation}
}
\begin{figure}
    \centering
    \includegraphics[width=1.\linewidth]{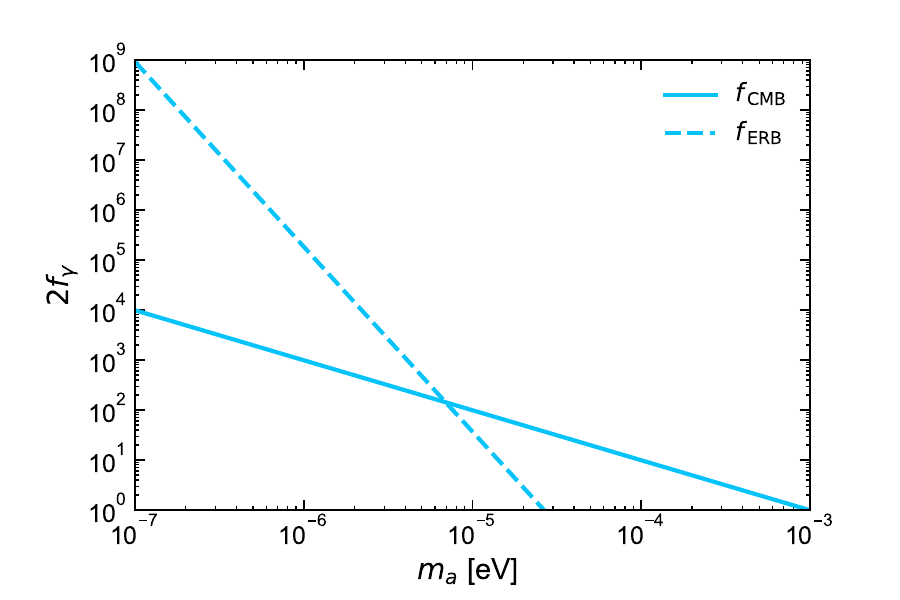}
    \caption{Stimulated emission factors $(2f_{\gamma})$ arising from the radio background at redshift $z = 0.05$. The solid blue line shows the contribution from the CMB, while the dashed blue line represents the extragalactic radio background contribution.}
    \label{fig:OccupationNumber}
\end{figure}

In Fig.~\ref{fig:OccupationNumber}, we present the stimulated emission factors ($2f_{\gamma}$) arising from the CMB (solid blue) and the ERB (dashed blue) as a function of the ALP DM mass at redshift $z=0.05$. 

The window function for the ALP DM decay signal, $W_{a}(\chi)$, characterizes the contribution to the observed intensity from redshift $z$. In natural units, it can be written as (see Appendix~\ref{App:Axion_WF} for a detailed derivation)
\begin{equation}
    W_{a}(\chi) = \frac{8\pi^{2}}{m_{a}^{3}}\frac{\Omega_{\rm c}\rho_{\rm cr}}{\tau_{a}}(1+z)^{2}\left(1 + 2f_{\gamma,{\rm CMB}} + 2f_{\gamma, {\rm ERB}}\right),
\end{equation}
where $\rho_{\rm cr} = 3H_{0}^{2}/8\pi G = 1.879h^{2}\times10^{-29}\,{\rm g\,cm^{-3}}$ is the critical density at present time. For the analysis of cross-correlations, with a given ALP dark matter mass around $\mu{\rm eV}$ scales, a specific redshift ranges corresponding to specific radio bands. To account for this, we multiply $W_{a}$ in redshift space by a redshift window function $W_{0}(z)$, defined as
\begin{equation}
    W_{0}(z) = \frac{\theta(z - z_{\rm min})\theta(z_{\rm max} - z)}{z_{\rm max} - z_{\rm min}},
\end{equation}
where $z_{\rm min}$ and $z_{\rm max}$ represent the lower and upper edges of the redshift bin under consideration, and $\theta$ is the Heaviside step function.

Regarding the spatial distribution, $g_{a}=\rho_{a}$, where we assume ALP DM follows a NFW density profile within halos
\begin{equation}
    \rho_{a} = \frac{\rho_{\rm s}}{r/r_{\rm s}\left(1 + r / r_{\rm s}\right)^{2}},
\end{equation}
where scale radius $r_{\rm s}(M,z) = R_{\rm vir}(M,z)/c_{\rm vir}(M,z)$ with halo virial radius $R_{\rm vir}(M,z)$ and concentration factor $c_{\rm vir}(M,z)$~\citep{Coe:2010xg}. The scale density can be expressed analytically as
\begin{equation}
    \rho_{\rm s}(M,z) = \frac{M}{4\pi r_{\rm s}^{3}}\left[\ln(1+c_{\rm vir}) - \frac{c_{\rm vir}}{1+c_{\rm vir}}\right]^{-1}.
\end{equation}

The Fourier transform of $\rho_{a}$ is then given by
\begin{equation}
\begin{aligned}
\tilde{\rho}_{a}(k|M,z) 
&= \int_{0}^{R_{\rm vir}} \mathrm{d}\vec{x}\rho_{a}(\vec{x}|M,z) e^{-i\vec{k}\cdot \vec{x}} \\
&= 4\pi r_{\rm s}^{3}\rho_{\rm s}
\Bigg\{
\cos (k r_{\rm s})
\Big[
\mathrm{Ci}\big((1+c_{\rm vir})k r_{\rm s}\big)
- \mathrm{Ci}(k r_{\rm s})
\Big] \\
&\quad
+ \sin (k r_{\rm s})
\Big[
\mathrm{Si}\big((1+c_{\rm vir})k r_{\rm s}\big)
- \mathrm{Si}(k r_{\rm s})
\Big] \\
& \quad- \frac{\sin (k r_{\rm s}) c_{\rm vir}}
{k r_{\rm s}(1+c_{\rm vir})},
\Bigg\}
\end{aligned}
\end{equation}
where $\mathrm{Ci}$ and $\mathrm{Si}$ denote the cosine and sine integral functions, respectively. The averaged cosmological ALP DM density is obtained by integrating the halo mass function over the halo population
\begin{equation}
    \left\langle\rho_{a}\right\rangle(z) = \int{\rm d}M \frac{{\rm d}n}{{\rm d}M}\int {\rm d}\vec{x}\rho_{a}(\vec{x}|M,z).
\end{equation}

Finally, by adopting the halo bias $b_{\rm h}$ following the Sheth-Tormen form \citep{Sheth:1999mn}, the resulting auto-power spectrum of the ALP DM density field relevant for the decay signal at $z=0.05$ is shown in Fig.~\ref{fig:ThreeD_PowerSpectrum} as the purple curve.

\subsection{Galaxy catalog}
In the local universe, Two Micron All Sky Survey (2MASS) provides a nearly exhaustive census of the galaxy distribution \citep{2006AJ....131.1163S}. Specifically, the 2MASS Redshift Survey (2MRS) \citep{2012ApJS..199...26H} contains spectroscopic redshifts for $N_{\rm g}=43182$ galaxies out to $z\sim 0.1$, covering an approximately sky fraction of $f_{\rm 2MRS}=0.877$~\cite{Ando:2017wff}. Following the methodology in Ref.~\cite{Ando:2017wff}, we adopt the normalized galaxy redshift distribution
\begin{equation}
    \frac{{\rm d}N_{\rm 2MRS}}{{\rm d}z} =\frac{\beta}{z_{0}\Gamma\left[(m+1) / \beta\right]}\left(\frac{z}{z_{0}}\right)^{m}\exp\left[-\left(\frac{z}{z_{0}}\right)^{\beta}\right],
\end{equation}
where parameters are set to $\beta = 1.64$, $z_{0} = 0.0266$, and $m=1.31$. The corresponding galaxy window function is then given by 
\begin{equation}
    W_{\rm g}(\chi) = \frac{{\rm d}N_{\rm 2MRS}}{{\rm d}z} \frac{H(z)}{c}.
\end{equation}

To relate the galaxy distribution to the underlying dark matter halos, we employ the Halo Occupation Distribution (HOD) framework. In this model, the spatial distribution of galaxies within a halo of mass $M$ is decomposed into contributions from a central galaxy and multiple satellite galaxies \cite{Ando:2014aoa, Ammazzalorso:2018evf}
\begin{equation}
\begin{aligned}
    g_{\rm g}(\vec{x}-\vec{x}'|M,z)  = &\, N_{\rm cen}(M)\delta_{\rm D}^{(3)}(\vec{x} - \vec{x}') \\
    +&\, N_{\rm sat}(M)\rho_{a}(\vec{x} - \vec{x}'|M, z)/M,
\end{aligned}
\end{equation}
where $\delta_{\rm D}^{(3)}$ denotes 3D Dirac delta function, assuming the central galaxy resides at the halo center. And the satellite galaxies are assumed to follow the dark matter density profile $\rho_{a}$.

The mean occupation numbers for central and satellite galaxies are parameterized as follows~\cite{Zheng:2004id, Ando:2014aoa,Ammazzalorso:2018evf}
\begin{equation}
\begin{aligned}
    N_{\rm cen} &= \frac{1}{2}\left[1 + {\rm erf}\left(\frac{\log_{10}M - \log_{10}M_{\rm min}}{\sigma_{\log_{10}M}}\right)\right], \\
    N_{\rm sat} & = 
    \begin{cases}
        \displaystyle\left(\frac{M - M_{0}}{M_{1}}\right)^{\alpha}, & M \geq M_0, \\[8pt]
        0, & M < M_0,
    \end{cases}
\end{aligned}
\end{equation}
where $\rm erf$ is the error function. We adopt the best-fit parameters for the 2MRS sample with $\log_{10}\left(M_{\rm min}/M_{\odot}\right) = 11.68$, $\sigma_{\log_{10}M}=0.15$, $\log_{10} \left(M_{0}/M_{\odot}\right) = 11.86$, $\log_{10} \left(M_{1}/M_{\odot}\right) = 13.0$, and $\alpha = 1.02$.

The averaged number density of galaxies at a given redshift $z$ is obtained by integrating the HOD over the halo mass function
\begin{equation}
    \langle g_{\rm g}\rangle(z) = \int {\rm d}M \frac{{\rm d}n}{{\rm d}M}\int{\rm d}\vec{x} g_{\rm g}(\vec{x}-\vec{x}'|M,z)
\end{equation}

In the Fourier domain, the spatial distribution of galaxies within a halo of mass $M$ is characterized by
\begin{equation}
    \tilde{g}_{\rm g}(k|M,z) = N_{\rm cen}(M) + \frac{N_{\rm sat}(M)}{M}\tilde{\rho}_{a}(k|M,z),
\end{equation}
where $\tilde{\rho}_{a}$ is the Fourier transform of dark matter density profile. 
Meanwhile the scale-independent galaxy bias $b_{\rm g}(z)$ is expressed as
\begin{equation}
    b_{\rm g}(z) = \int {\rm d}M \frac{{\rm d}n}{{\rm d}M} b_{\rm h}(M,z)\frac{N_{\rm g}(M)}{\langle g_{\rm g} \rangle(z)},
\end{equation}
where $N_{\rm g}(M) = N_{\rm cen}(M) + N_{\rm sat}(M)$.

Using these ingredients, we compute the galaxy auto-power spectrum and the galaxy-ALP DM decay cross-power spectrum. In Fig.~\ref{fig:ThreeD_PowerSpectrum}, we present the galaxy-ALP cross-power spectrum (solid black) and the galaxy auto-power spectrum (solid green) evaluated at $z=0.05$. For reference, the linear matter power spectrum at the same redshift is indicated by the dotted gray line.

\begin{figure}
    \centering
    \includegraphics[width=1.\linewidth]{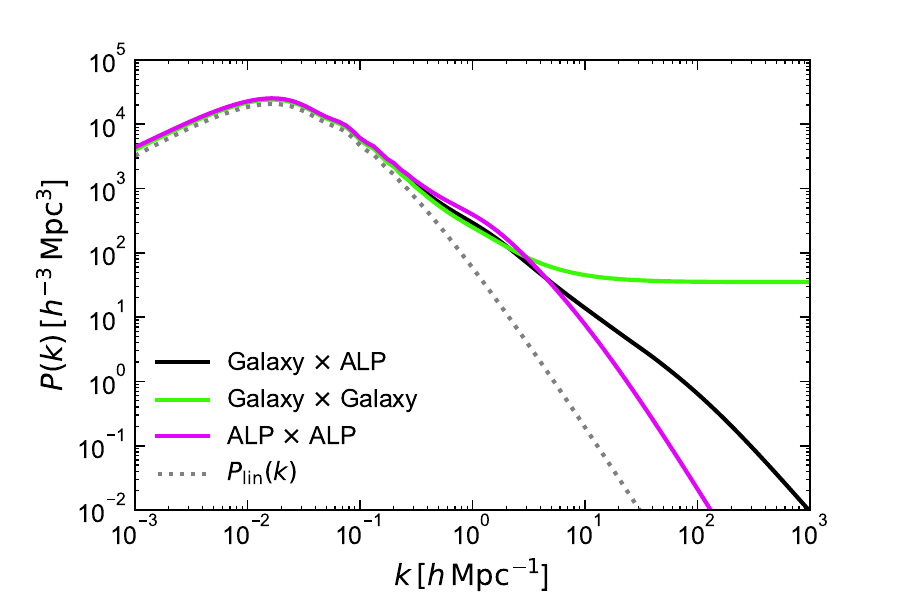}
    \caption{Power spectra at redshift $z=0.05$ for the cross-correlation between galaxies and ALP DM decay (solid black), galaxy auto-correlation (solid green), and ALP DM auto-correlation (solid purple).The linear matter power spectrum at the same redshift is shown by the dotted gray line.}
    \label{fig:ThreeD_PowerSpectrum}
\end{figure}

\section{Sensitivity Forecast and Discussion}\label{Sec:ResultDiscussion}

The core of our detection strategy relies on the redshift-frequency mapping inherent to ALP DM decay. For a given ALP DM mass $m_{a}$, the monochromatic emission is observed at a frequency $\nu_{\rm obs} = m_{a}/[4\pi (1+z)]$ that falls within the detection bandwidth of SKA. Since the 2MRS catalog provides a robust tracer of the LSS up to $z\approx 0.1$, any potential ALP DM decay signal must be spatially correlated with these galaxies. Consequently, scanning through the SKA frequency bands is equivalent to searching for ALP dark matter candidates across a wide mass range.

Using the window functions and power spectra derived in the previous sections, the APS is computed via the Limber approximation (see Eq.~(\ref{APS})). To assess the detectability, we define the variance of the cross-correlation estimator $\left(\Delta C_{\ell}^{a{\rm g}}\right)^{2}$ as
\begin{equation}
    \left(\Delta C_{\ell}^{a{\rm g}}\right)^{2} = \frac{f_{\rm sky}^{-1}}{(2\ell + 1)}\left[\left(C_{\ell}^{a{\rm g}}\right)^{2} + \left(C_{\ell}^{aa} + \frac{C_{\rm N}}{B_{\ell}^{2}}\right)\left(C_{\ell}^{\rm gg} + C_{\rm Ng}\right)\right],
\end{equation}
where $f_{\rm sky} = \min\{f_{\rm 2MRS}, f_{\rm SKA}\}$ denotes the effective sky coverage fraction, with $f_{\rm SKA}$ the sky coverage fraction of SKA, while $C_{\ell}^{aa}$ and $C_{\ell}^{\rm gg}$ represent the auto-correlation APS for ALP decay and galaxies, respectively.

The term $C_{\rm N}$ accounts for the radio telescope's thermal noise, while $B_{\ell}$ is the beam window function that characterizes the angular resolution limits. The galaxy shot noise is given by $C_{\rm Ng}=4\pi f_{2\rm MRS}/N_{\rm g}$. Notably, we do not apply a beam correction to the galaxy term, as the angular resolution of spectroscopic galaxy catalogs far exceeds the angular scales relevant to this analysis.

Fig.~\ref{fig:APS} illustrates the resulting APS for a benchmark ALP DM candidate with $m_{a}=4.136\,\mu{\rm eV}$ and $g_{a\gamma\gamma} = 10^{-9}\,{\rm GeV}^{-1}$. To facilitate a comparison of their respective scale dependencies, we present these spectra on the same axes despite their differing units. The cross-correlation APS between galaxies and ALP DM decay is shown by the black line (in units of $[\mu{\rm K}]$), while the purple line represents the ALP DM decay auto-APS (in units of $[\mu{\rm K}]^{2}$). Additionally, the dimensionless galaxy auto-correlation APS is indicated by the green line.

\begin{figure}
    \centering
    \includegraphics[width=1.\linewidth]{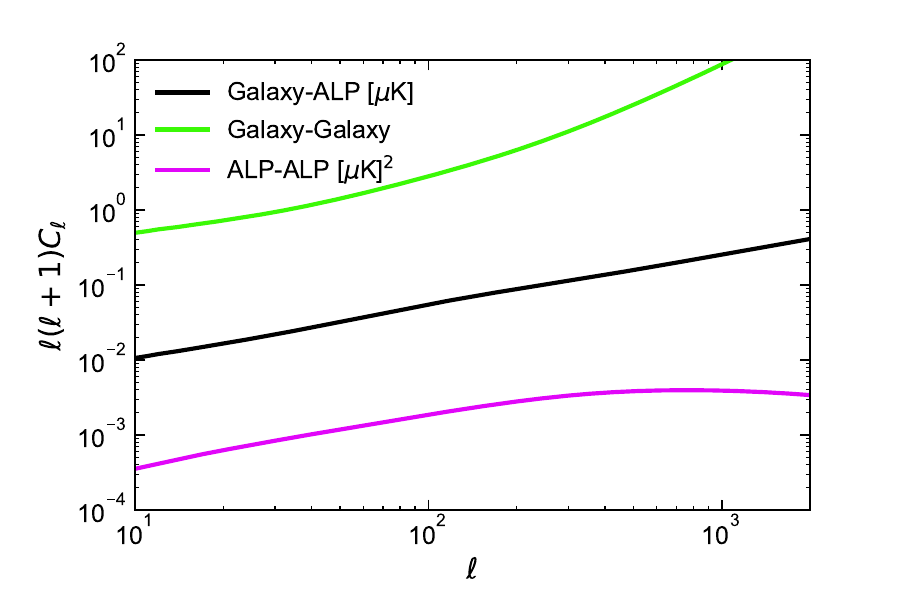}
    \caption{Angular power spectrum $C_{\ell}$ integrated over the redshift range $z=0$ to $z=0.1$. Shown are the galaxy–ALP DM decay cross-correlation (black; in units of $[\mu{\rm K}]$), the galaxy auto-correlation (green; dimensionless), and the ALP DM auto-correlation (purple; in units of $[\mu{\rm K}]^{2}$). A benchmark ALP DM candidate is assumed with $m_{a}=4.136\,\mu{\rm eV}$ and $g_{a\gamma\gamma} = 10^{-9}\,{\rm GeV^{-1}}$.}
    \label{fig:APS}
\end{figure}

We forecast the sensitivity based on the projected performance of SKA2 in single-dish mode. For this analysis, we assume a frequency-independent instrumental response across the broad range from $50\,{\rm MHz}$ to $15\,{\rm GHz}$. The angular response of the instrument is modeled as a Gaussian beam, yielding\cite{Battye:2012tg}
\begin{equation}
    B_{\ell} = \exp \left[-\frac{\ell^{2}}{2}\left(\frac{1.22}{\sqrt{8\ln2}}\frac{\lambda_{\rm obs}}{D}\right)^{2}\right],
\end{equation}
where $\lambda_{\rm obs}$ is the observed wavelength and $D$ is the dish diameter. The thermal noise power spectrum for a single-dish scan is modeled as \cite{Camera:2013kpa, Pinetti:2019ztr}
\begin{equation}
    C_{\rm N} = \frac{T_{\rm sys}^{2}S}{N_{\rm d}t_{\rm obs}\Delta\nu N_{\rm b} N_{\rm pol} \eta^{2}},
\end{equation}
where $T_{\rm sys} = 30 + 60(300\,{\rm MHz} / \nu)^{2.55}\,{\rm K}$ accounts for the combined frequency-dependent sky temperature and receiver noise. Here, $S$ is the survey area in steradians, $N_{\rm d}$ is the number of dishes, $t_{\rm obs}$ is the integration time, $\Delta \nu$ is the frequency bandwidth corresponding to the redshift bin considered, and $N_{\rm pol}=2$ is the number of polarization states. We incorporate $N_{\rm b}$ to account for simultaneous beams facilitated by Phased Array Feeds (PAFs), which significantly enhance survey speed. The efficiency $\eta$ is assumed to be unity for this ideal sensitivity forecast. To quantify the forecasting performance of the SKA2, we adopt the technical specifications from Ref.~\cite{Pinetti:2019ztr}, as summarized in Table~\ref{tab:SKA2}.

\begin{table}[t]
    \centering
    \caption{Technical specifications for the SKA2, adopted from Ref.~\cite{Pinetti:2019ztr}.}
    \label{tab:SKA2}
    \addtolength{\tabcolsep}{1pt} 
    \begin{tabular}{c ccccc}
        \toprule
        \toprule
        Survey Area $S$ & $t_{\rm obs}$ & $N_{\rm d}$ & $D_{\rm dish}$ & $N_{\rm b}$ & $f_{\rm SKA}$ \\
        $[{\rm deg^{2}}]$ & $[10^{3}\, {\rm hr}]$ & & $[{\rm m}]$ & & \\
        \midrule
        30,000 & 10 & 2,000 & 14.5 & 36 & 0.72 \\
        \bottomrule
        \bottomrule
    \end{tabular}
\end{table}

The total signal-to-noise ratio (SNR) for the cross-correlation signal is obtained by summing over all available multipoles
\begin{equation}
    {\rm SNR}^{2} = \sum_{\ell}\left(\frac{C_{\ell}^{a{\rm g}}}{\Delta C_{\ell}^{a{\rm g}}}\right)^{2}.
\end{equation}

We evaluate the sensitivity of this cross-correlation approach by testing the ALP DM signal against the null hypothesis (i.e., a universe with no ALP DM decay). For each ALP DM mass candidate $m_{a}$, we derive the upper limit on the ALP-photon coupling parameter $g_{a\gamma\gamma}$ at the $95\%$ confidence level (C.L.) by requiring a threshold of $\rm SNR = 2$. The resulting sensitivity curves for $g_{a\gamma\gamma}$ as a function of $m_{a}$ are presented in Fig.~\ref{fig:UpperLimits}, where we compare our results with existing constraints from the CAST~\citep{CAST:2017uph}.

Our analysis highlights the transition between different stimulated emission regimes. For ALP masses below $\mathcal{O}(10\,\mu{\rm eV})$, the stimulated decay is primarily driven by the ERB. We model the ERB using a bottom-up approach based on the SFR–$L_\nu$ relation and the SHMR. While these empirical relations are well calibrated in the low-redshift Universe, their extrapolation to higher redshifts introduces a degree of model uncertainty that may affect the inferred ERB intensity.{ Anisotropies in the ERB have been reported~\cite{Offringa:2021rwp}, although their origin remains uncertain. In this work, we consider only the mean (isotropic) ERB component. Fluctuations in the ERB could, in principle, modulate the ALP DM signal and contribute to the angular power spectrum by introducing additional three-dimensional power spectrum components involving the ERB field. Using our ERB model, we estimate their impact and find that these contributions are strongly suppressed relative to the local dark matter density fluctuations. This suppression arises from the line-of-sight integration inherent in the ERB construction. Therefore, contributions from ERB fluctuations can be safely neglected in the present analysis.} As $m_a$ increases beyond $10\,\mu{\rm eV}$, the CMB becomes the dominant photon field.

Another important consideration in intensity mapping is the presence of interlopers, namely astrophysical emission lines that fall within the same observed frequency bands. For instance, the HI 21cm line originating from redshifts $z=0$–$0.1$ occupies the frequency range $1.29$–$1.42\,{\rm GHz}$. In searches for ALP-induced signals within this band, HI emission may act as a contaminant, requiring careful modeling of foreground contributions in future analyses.

\begin{figure}
    \centering
    \includegraphics[width=1.\linewidth]{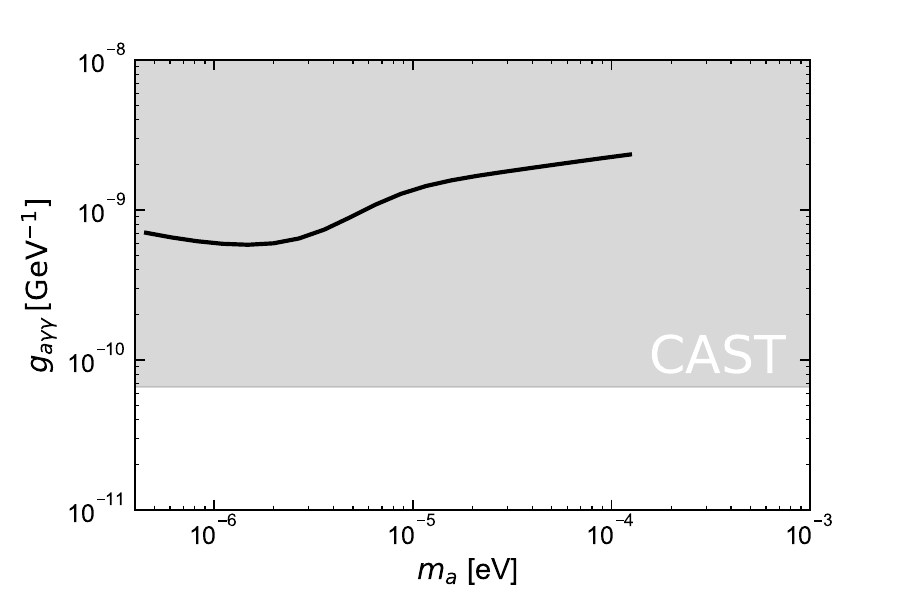}
    \caption{Projected $95\%$ C.L. upper limits on the ALP-photon coupling $g_{a\gamma\gamma}$ as a function of the ALP DM mass $m_{a}$. The sensitivity estimates are compared with existing constraints from the CAST helioscope \citep{CAST:2017uph}.}
    \label{fig:UpperLimits}
\end{figure}

\section{Conclusion}\label{Sec:Conclusion}
ALPs have emerged as some of the most compelling candidates for cold dark matter, garnering significant theoretical and observational interest. The fundamental coupling between ALPs and the electromagnetic field facilitates two primary detection channels, the spontaneous or stimulated decay into photon pairs and the Primakoff conversion into a single photon in the presence of an external magnetic field. These mechanisms provide the theoretical bedrock for diverse experimental searches. In this work, we have presented a comprehensive sensitivity forecast for detecting ALP dark matter by cross-correlating radio intensity mapping with the 2MASS Redshift Survey within the local volume ($z\leq 0.1$). By constructing a bottom-up model for the extragalactic radio background and accounting for stimulated emission effects, we have established a robust framework to evaluate ALP-induced line emissions on cosmological scales. Our analysis demonstrates that the angular cross-power spectrum between the projected SKA2 intensity maps and large-scale galaxy distributions can effectively disentangle faint ALP-induced signals from dominant astrophysical foregrounds. The projected sensitivity for the ALP-photon coupling, $g_{a\gamma\gamma}$, indicates that this technique {serves as a promising complementary probe} within the $\mu{\rm eV}$ mass regime. {While current forecasts may not surpass leading helioscope bounds, this work provides a vital proof-of-concept for exploring the ALP dark matter parameter space through large-scale structure correlations.}

Looking forward, several avenues remain to refine and extend this methodology. The fidelity of our ERB modeling can be further improved by incorporating more sophisticated galaxy evolution prescriptions and high-redshift radio luminosity functions. Furthermore, extending this cross-correlation framework to include multi-tracer analyses, such as cross-correlations with weak gravitational lensing maps, could offer a more direct probe of the underlying dark matter distribution.

\section{Acknowledgments}
The financial assistance of the South African Radio Astronomy Observatory (SARAO) towards this research is hereby acknowledged (\url{www.sarao.ac.za}).

\appendix
\section{ALP DM Decay Window Function}\label{App:Axion_WF}
Radio intensity mapping measures the sky brightness in units of temperature, which is related to the specific intensity $I_{\nu}$ through the Rayleigh-Jeans approximation
\begin{equation}
T_{\rm b} = \frac{c^{2}}{2k_{\rm B}\nu^{2}}I_{\nu}.
\end{equation}

In the following derivation, we adopt natural units where $c=\hbar=k_{\rm B}=1$. For ALP dark matter decay, the monochromatic emission coefficient at frequency $\nu_{\rm e}$ and redshift $z$ is given by
\begin{equation}
    j_{\nu_{\rm e}}(z) = 2\frac{\rho_{a}(z)}{4\pi m_{a}\tau_{a}}E_{\gamma}\delta_{\rm D}\left(\nu(1+z) - \nu_{\rm e}\right),
\end{equation}
where $\rho_{a}(z)$ is the ALP energy density, $\tau_{a}$ is the decay lifetime, $E_{\gamma} = m_{a}/2$ is the energy of each emitted photon, and $\delta_{\rm D}$ denotes Dirac delta function. The factor of $2$ accounts for the two photons produced per decay event. The rest-frame emission frequency is $\nu_{\rm e}=m_{a}/4\pi$.

Tanking in account the expansion of the universe and the invariance of $I_{\nu}/\nu^{3}$ along photon geodesics, the observed specific intensity from a comoving distance element ${\rm d}\chi$ is 
\begin{equation}
\begin{aligned}
    {\rm d}I_{\nu} & = \frac{\nu^{3}}{\nu_{\rm e}^{3}}{\rm d}I_{\nu_{\rm e}} \\
    & = \frac{1}{(1+z)^{3}}{\rm d}I_{\nu_{\rm e}} \\
    & = \frac{j_{\nu_{\rm e}}}{(1+z)^{3}}{\rm d}s \\
    & = \frac{j_{\nu_{\rm e}}}{(1+z)^{4}}{\rm d}\chi,
\end{aligned}
\end{equation}
where ${\rm d}s = (1+z)^{-1}{\rm d}\chi$ is the proper distance element. Combining the above relations, we define the ALP DM decay window function, namely $W_{a}(\chi, \nu)$, as
\begin{equation}
\begin{aligned}
    W_{a}(\chi, \nu) & = \frac{\partial T}{\partial\chi} \\
    & = \frac{1}{2\nu^{2}} \frac{j_{\nu_{\rm e}}}{(1+z)^{4}} \\
    & = \frac{1}{8\pi \nu^{2}} \frac{\Omega_{\rm c}\rho_{\rm cr}}{\tau_{a}(1+z)}\delta_{\rm D}\left(\nu(1+z)-\nu_{\rm e}\right) \\
    & = \frac{1}{8\pi \nu^{3}} \frac{\Omega_{\rm c}\rho_{\rm cr}}{\tau_{a}(1+z)}\delta_{\rm D}\left((1+z)-\nu_{\rm e}/\nu\right)\\
    & = \frac{8\pi^{2}}{m_{a}^{3}}\frac{\Omega_{\rm c}\rho_{\rm cr}}{\tau_{a}}(1+z)^{2}\delta_{\rm D}\left(z - z_{\rm e}\right),
\end{aligned}
\end{equation}
where $\nu = \nu_{\rm e}/(1+z) = m_{a}/4\pi(1+z)$ is observed photon frequency. The above expression shows that, for a fixed ALP dark matter mass, decay occurring at redshift $z_{\rm e}$ produces emission observed at frequency $\nu$.

In practical observations, a radio telescope operates over a finite frequency bin $[\nu_{\rm min}, \nu_{\rm max}]$. To model the signal within specific frequency channels, the Dirac delta function is replaced by a normalized redshift window function $W_{0}(z)$, defined as
\begin{equation}
    W_{0}(z) = \frac{\theta(z - z_{\rm min})\theta(z_{\rm max} - z)}{z_{\rm max} - z_{\rm min}},
\end{equation}
where $z_{\rm min}$ and $z_{\rm max}$ are the redshift boundaries corresponding to the frequency interval, and $\theta$ is the Heaviside step function.

Furthermore, considering the stimulated decay process induced by background photons, the ALP dark matter decay window function can be written as
\begin{equation}
    W_{a}(\chi, \nu) = \frac{8\pi^{2}}{m_{a}^{3}}\frac{\Omega_{\rm c}\rho_{\rm cr}}{\tau_{a}}(1+z)^{2}(1 + 2f_{\gamma}),
\end{equation}
where the factor $(1+2f_{\gamma})$ accounts for the combined contribution from spontaneous and stimulated decay processes.

\clearpage
\bibliographystyle{unsrt}
\bibliography{sample}

\end{document}